\documentclass[reprint,
superscriptaddress,amsmath,amssymb,aps, longbibliography
]{revtex4-2}

\usepackage{graphicx}
\usepackage{dcolumn}
\usepackage{bm}
\usepackage{siunitx}
\usepackage[noend]{algorithmic}
\usepackage{algorithm}
\usepackage[skins]{tcolorbox}
\usepackage{setspace}
\usepackage{blkarray}

\usepackage{xcolor}
\usepackage{soul}
\usepackage{systeme}
\usepackage{footmisc}

\usepackage{scalerel} 

\makeatletter
\def\maketag@@@#1{\hbox{\m@th\normalfont\normalsize#1}}
\makeatother

\newcommand{\sgn}{\text{sgn}}
\newcommand{\blue}[1]{\textcolor{black}{#1}}
\newcommand{\cyan}[1]{\textcolor{black}{#1}}

\newcommand{\tr}{\text{tr}}

\usepackage{booktabs}
\usepackage{hyperref}
\usepackage[capitalize]{cleveref}
\usepackage[utf8]{inputenc}

\usepackage[lining,semibold]{libertine} 
\usepackage{amsthm}
\usepackage[libertine, cmintegrals, bigdelims, vvarbb]{newtxmath}

\usepackage{amsmath}
\usepackage{amsfonts}
\usepackage{mathrsfs}
\usepackage{gensymb}
\usepackage{bbm}
\usepackage{dsfont}

\usepackage{pgfplots}
\usepgfplotslibrary{colormaps}

\usepackage{kbordermatrix}
\renewcommand{\kbldelim}{(}
\renewcommand{\kbrdelim}{)}

\usepackage{chemformula}
\usepackage[caption=false]{subfig}

\usepackage{soul}
\usepackage{xcolor}

\theoremstyle{definition}

\usepackage{scalerel} 

\definecolor{webgreen}{rgb}{0,.5,0}
\definecolor{webbrown}{rgb}{.6,0,0}
\definecolor{grigio}{rgb}{.85,.85,.85} 
\definecolor{RoyalBlue}{rgb}{0.0, 0.14, 0.4}
\definecolor{skyblue1}{rgb}{0.45,0.62,0.81}
\definecolor{skyblue2}{rgb}{0.2,0.39,0.64}
\definecolor{skyblue3}{rgb}{0.13,0.29,0.53}
\definecolor{scarlet1}{rgb}{0.93,0.16,0.16}
\definecolor{scarlet2}{rgb}{0.8,0,0}
\definecolor{scarlet3}{rgb}{0.64,0,0}

\definecolor{g}{gray}{0.50}

\hypersetup{%
    colorlinks=true, linktocpage=true, pdfstartpage=1, pdfstartview=FitV,%
    breaklinks=true, pdfpagemode=UseNone, pageanchor=true, pdfpagemode=UseOutlines,%
    plainpages=false, bookmarksnumbered, bookmarksopen=true, bookmarksopenlevel=1,%
    hypertexnames=true, pdfhighlight=/O,
    urlcolor=webbrown, linkcolor=RoyalBlue, citecolor=webgreen, 
    pdftitle={},%
    pdfauthor={Timur Aslyamov},%
    pdfsubject={},%
    pdfkeywords={},%
    pdfcreator={pdfLaTeX},%
    pdfproducer={LaTeX REVTeX}%
}

\begin{document}
\title{General Theory of Static Response for Markov Jump Processes}

\author{Timur Aslyamov}
\email{timur.aslyamov@uni.lu}

\author{Massimiliano Esposito}
\email{massimiliano.esposito@uni.lu}
\affiliation{Department of Physics and Materials
Science, University of Luxembourg, L-1511 Luxembourg City, Luxembourg}

\date{\today}

\begin{abstract}
We consider Markov jump processes on a graph described by a rate matrix that depends on various control parameters. 
We derive explicit expressions for the static responses of edge currents and steady-state probabilities. 
We show that they are constrained by the graph topology (i.e. the incidence matrix) by deriving response relations (i.e. linear constraints linking the different responses) and topology-dependent bounds. 
For unicyclic networks, all scaled current 
\blue{responses}
are between zero and one and must sum to one.
Applying these results to stochastic thermodynamics, we derive explicit expressions for the static response of fundamental currents (which carry the full dissipation) to fundamental thermodynamic forces (which drive the system away from equilibrium). 
\end{abstract}
\maketitle


\textit{Introduction.---}Nonequilibrium steady state (NESS) of Markov jump processes describe a plethora of phenomena \cite{zhang2012stochastic} and understanding their response to external perturbations has crucial implications \cite{ge2012stochastic} across fields, such as biology \cite{owen2023size,cornish2013fundamentals,fell1997understanding,murugan2014discriminatory}, nanoelectronics \cite{freitas2021stochastic,moreira2023stochastic}, and deep learning \cite{sohl2015deep}. 
When the Markov jump process is produced by thermal noise, near equilibrium, the response is simple and characterized by the dissipation-fluctuation relation (DFR) \cite{kubo1966fluctuation,govern2014energy}. But far-from-equilibrium or for non-thermal processes, the response is significantly more involved; see Refs.~\cite{agarwal1972fluctuation,seifert2010fluctuation,prost2009generalized,altaner2016fluctuation,baiesi2009fluctuations,baiesi2013update,hatano2001steady,falasco2019negative,shiraishi2022time,dechant2020fluctuation,szabo2007evolutionary, chun2023trade, gao2024thermodynamic}.
Recently, for the static response, exact results~\cite{aslyamov2024nonequilibrium,de2023fermionic, floyd2024learning} and tight bounds~\cite{owen2020universal,gabriela2023topologically} have been derived assuming Arrhenius-like rates.
Moreover, the bounds~\cite{aslyamov2024nonequilibrium} only hold for local responses, i.e. when the perturbation and the observable are assigned to the same transition.   

In this Letter, we build a general static response theory for any Markov jump processes, which describes both local and nonlocal responses for arbitrary parameterizations of the rate matrix. We identify a broad class of parameterizations that produce two types of linear constraints, which we call the Summation Response Relation (SRR) and Cycle Response Relation (CRR). 
The SRR restricts the responses of the edge flux and probability. The form of such constraints does not depend on the topology of the incidence matrix, but the values of the responses involved strongly depend on it. The CRR limits the sum of local responses by the number of fundamental (Schnakenberg \cite{schnakenberg1976network}) cycles, which are essential topological characteristics.  Moreover, the topology defines which static responses among all combinatorial configurations have universal (remarkably simple) bounds.
In unicyclic networks, all responses are bounded; for multicyclic systems, our approach identifies bounded and unbounded responses. In concrete examples considered, the sizes of bounded and unbounded sets are comparable. 
Finally, for Markov jump processes describing a system in contact with thermal reservoirs (i.e. stochastic thermodynamics), we derive an explicit expression for the static response of fundamental currents to fundamental thermodynamic forces. The former characterize the full dissipation and the latter drive the system out-of-equilibrium.

\textit{Setup.}---We consider a directed graph $\mathcal{G}$ with $N$ nodes and $N_{e}$ edges and a Markov jump process over the discrete set $\mathcal{S}$ of the $N$ states corresponding to the nodes. Then, the edges $e\in\mathcal{E}$ of $\mathcal{G}$ define possible transitions with the probability rates encoded in the rate matrix $\mathbb{W}/\tau$. In this description, the jump from $n$ to $m$ is the edge (arrow) $+e$ with the source $s(+e)=n$ and the tip $t(+e)=m$. For the reverse transition, we have $-e$ with $s(-e)=m$ and $t(-e)=n$. Choosing $\tau=1$ the non-diagonal elements $W_{nm}=W_{e}\geq 0$ become the probabilities per unit of time assigned to the edges $e$. 
We assume that the matrix $\mathbb{W}$ is irreducible \cite{van1992stochastic} and that all transitions are reversible, i.e. $W_{e}\neq 0$ only if $W_{-e}\neq 0$.
With the property of diagonal elements $W_{ii}=-\sum_{j\neq i}W_{ij}$ the described system always exhibits a unique steady-state probability $\boldsymbol{\pi}=(\pi_1,\dots,\pi_N)^\intercal$ that satisfies 
\begin{align}
    \label{eq:master_eq}
        \mathbb{W}\cdot\boldsymbol{\pi}&=\boldsymbol{0}\,,
\end{align}
with the normalization $\sum_{i=1}^N{\pi_i}=1$. We define the transition current along the edge $e$ as $j_{e} \equiv W_{+e}\pi_{s(+e)}-W_{-e}\pi_{s(-e)}$. This definition has a matrix form $\boldsymbol{j}=\mathbb{\Gamma}\boldsymbol{\pi}$,
where the matrix $\mathbb{\Gamma}$ has elements $\Gamma_{ei} \equiv W_{+e}\delta_{is(+e)}-W_{-e}\delta_{is(-e)}$ with the Kronecker symbol $\delta$.
In analogy to linear chemical reaction networks~\cite{gunawardena2012linear,srinivas2023characterizing,aslyamov2023nonideal}, the rate matrix can be decomposed as $\mathbb{W}=\mathbb{S}\mathbb{\Gamma}$, where $\mathbb{S}$ is the incidence matrix of the directed graph $\mathcal{G}$ with the elements $S_{ie} \equiv \delta_{is(-e)}-\delta_{is(+e)}$.

We introduce a parameterization $\mathbb{W}(\boldsymbol{p})$ using the vector $\boldsymbol{p}=(\dots,p,\dots)^\intercal$ made of $N_p$ control parameters $p\in \boldsymbol{p}$. 
Thus, the steady-state condition \eqref{eq:master_eq} can be written as 
\begin{align}
    \label{eq:matS_steady-state}
    \mathbb{S}\cdot\boldsymbol{j}(\boldsymbol{p})=\boldsymbol{0}\,,
\end{align}
where $\boldsymbol{j}(\boldsymbol{p})\equiv \boldsymbol{j}(\boldsymbol{p},\boldsymbol{\pi}(\boldsymbol{p}))$ depends on $\boldsymbol{p}$ both explicitly due to $\mathbb{\Gamma}(\boldsymbol{p})$ and implicitly due to $\boldsymbol{\pi}(\boldsymbol{p})$. The central objects of this work are the static responses of a quantity $q(\boldsymbol{p})$ (e.g. a probability $\boldsymbol{\pi}$ or a current $\boldsymbol{j}$) to perturbations of the parameters $p$, i.e. the elements of the vector $\nabla_{\boldsymbol{p}} q(\boldsymbol{p})=(\dots,d_p q ,\dots)^\intercal$. 
\blue{Our goal will be to obtain \textit{explicit} relations for these responses enabling us to find relations among them.
Many recent works considered edge perturbations where each element of $\boldsymbol{p}$ acts on a rate associated with a given edge \cite{aslyamov2024nonequilibrium,harunari2024mutual,zheng2024information,floyd2024learning, de2023fermionic} . But often, perturbing physical quantities implies acting on rates associated with many edges.
In this letter, after an example, we go from generic perturbations to more specific ones, illustrating our findings at every stage using a physical model.} 

\textit{Homogeneous parameterization.---}We first provide a simple illustration of how a given parameterization can give rise to nontrivial relations amongst static responses, let us consider the parameterization $\boldsymbol{h}=\blue{(\dots,h_p,\dots)^\intercal}$ of $\mathbb{W}(\boldsymbol{h})$ such that:
\begin{align}
\label{eq:homogen-1}
\mathbb{W}(\alpha\boldsymbol{h})&=\alpha^k \mathbb{W}(\boldsymbol{h})\,,
\end{align}
where $k$ is the positive order of the homogeneous function. The fact that $\mathbb{W}(\alpha\boldsymbol{h})\boldsymbol{\pi}(\alpha\boldsymbol{h})=\alpha^k\mathbb{W}(\boldsymbol{h})\boldsymbol{\pi}(\alpha\boldsymbol{h})=0$ implies that $\boldsymbol{\pi}(\boldsymbol{h})=\boldsymbol{\pi}(\alpha\boldsymbol{h})$ since the solution of \cref{eq:master_eq} is unique; $\boldsymbol{\pi}(\boldsymbol{h})$ is therefore an homogeneous function of order zero of $\boldsymbol{h}$ which implies the linear relation (Euler's theorem for $k=0$) 
\begin{align}
\label{eq:relation-1-homog}
    \sum_{p} h_p \frac{d\boldsymbol{\pi}}{d h_p}&= 0\,.
\end{align}
This in turn implies that the current is a homogeneous function $\boldsymbol{j}(\alpha\boldsymbol{h},\boldsymbol{\pi}(\alpha\boldsymbol{h}))=\alpha^k\boldsymbol{j}(\boldsymbol{h},\boldsymbol{\pi}(\boldsymbol{h}))$, which is equivalent to:
\begin{align}
\label{eq:relation-2-homog}
 \sum_{p} h_p \frac{d\boldsymbol{j}}{d h_p}&=k \boldsymbol{j}\,.
\end{align}
\Cref{eq:relation-1-homog,eq:relation-2-homog} 
are known in metabolic control analysis as summation theorems~\cite{heinrich1974linear,kacser1995control,reder1988metabolic,heinrich2012regulation}. In that context, enzyme concentrations play the role of the homogeneous parameters. 

\textit{Matrix approach to static response.---} We now turn to arbitrary parameterizations. Our strategy is to use $\sum_{i=1}^{N}\pi_i=1$, and thus $d_{p}\pi_N=-\sum_{k=1}^{N-1} d_p\pi_k$, to arrive at $N-1$ independent equations for others $d_p\pi_k $ with $k\in \Hat{\mathcal{S}} \equiv \mathcal{S}\setminus\{N\}$. To solve this linear problem, we introduce the matrix $\mathbb{K} \equiv \Hat{\mathbb{S}}\Hat{\mathbb{\Gamma}}$, where $\Hat{\mathbb{S}} \equiv [S_{ke}]_{\{k,e\}}$ and $\Hat{\mathbb{\Gamma}} \equiv [\Gamma_{ek}-\Gamma_{eN}]_{\{e,k\}}$ are reduced matrices with $k\in\Hat{\mathcal{S}}$ and $e\in\mathcal{E}$.  
In Sec. A of 
\cite{noteSM} 
we prove that the matrix $\mathbb{K}$ is invertible [see Eq.~A5], and that the probability and current response matrices read:
\begin{subequations}
\label{eq:response-matrix}
    \begin{align}
    \label{eq:response-NESS-matrix}
    \mathbb{R}^\pi&
    \equiv [d_p \pi_i]_{\{i, p\}}=
    -
    \begin{pmatrix}
        \mathbb{K}^{-1}\Hat{\mathbb{S}}\mathbb{J}\\
        -\sum_{k} (\mathbb{K}^{-1}\Hat{\mathbb{S}}\mathbb{J})^\intercal_k
    \end{pmatrix}\,,\\
    \label{eq:response-current-matrix}
 \mathbb{R}^j&\equiv [d_p j_e]_{\{e, p\}}
 =\mathbb{P}\mathbb{J}\,.
    \end{align}
\end{subequations}
Here, $\sum_k(\mathbb{K}^{-1}\Hat{\mathbb{S}}\mathbb{J})^\intercal_k$ denotes the row that is the sum of all rows of the matrix $\mathbb{K}^{-1}\Hat{\mathbb{S}}\mathbb{J}$; $\{i, p\}$ (resp. $\{e, p\}$) denote the sets of indexes $i\in \mathcal{S}$ (resp. $e\in \mathcal{E}$) for rows and $p\in\boldsymbol{p}$ for columns; $\mathbb{J}\equiv[\partial_p j_e]_{\{e,p\}}$ is the steady state Jacobian
\begin{align}
\label{eq:matJ}
    \mathbb{J}
    =[\pi_{s(+e)}\partial_p W_e(\boldsymbol{p})-\pi_{s(-e)}\partial_p W_{-e}(\boldsymbol{p})]_{\{e,p\}} 
    \;,
\end{align}
and the matrix $\mathbb{P}=[P_{ee'}]_{\{e,e'\}}$ is defined as
\begin{align}
\label{eq:matP}
    \mathbb{P}&\equiv\Big[\delta_{ee'}-\sum_{x,x'\in\Hat{\mathcal{S}}}\Hat{\Gamma}_{ex}(\mathbb{K}^{-1})_{xx'}S_{x'e'}\Big]_{\{e,e'\}}=\mathbb{I}-\Hat{\mathbb{\Gamma}}(\Hat{\mathbb{S}}\Hat{\mathbb{\Gamma}})^{-1}\Hat{\mathbb{S}}\,,
\end{align}
with $\mathbb{I}$ denoting the identity matrix. Matrix $\mathbb{P}$ is idempotent [$\mathbb{P}^2=\mathbb{P}$] and is known as an oblique projection matrix \cite{brust2020comp}.
Since $\Hat{\mathbb{S}}\mathbb{P}=0$, we define $\mathbb{B}$ via
\begin{align}
\label{eq:matP-CB-form}
\mathbb{P}& \equiv \Big[\sum_{\gamma\in\mathcal{C}}c^\gamma_e B_{\gamma e'}\Big]_{\{e,e'\}}=\mathbb{C}\mathbb{B}\,,
\end{align}
where $\mathbb{C}=(\dots,\boldsymbol{c}_\gamma,\dots)$ is the matrix of the fundamental cycles $\boldsymbol{c}^\gamma$ defined as the right null vectors of the incidence matrix $\mathbb{S}\boldsymbol{c}^\gamma=\boldsymbol{0}$ ($\mathbb{S}$ and $\Hat{\mathbb{S}}$ share the same $N_c$ cycles). 
\Cref{eq:response-matrix} is crucial in what follows. \blue{It contains explicit expressions for the responses of all edges to arbitrary perturbations and contains information on how they are related to each other.}

\textit{Response relations.---}For a vector $\boldsymbol{p}$ such that 
the matrix $\mathbb{J}$ in \cref{eq:matJ} is full row rank ($\text{rk}\mathbb{J}=N_{e}$, i.e. $N_p\geq N_e$), we can always find a right invertible matrix $\mathbb{J}^+$ such that $\mathbb{J}\mathbb{J}^+=\mathbb{I}$:
\begin{align}
    \label{eq:right-invertible}
    \mathbb{J}^+\equiv\mathbb{J}^\intercal(\mathbb{J}\mathbb{J}^\intercal)^{-1}\,.
\end{align}
Multiplying both sides of \cref{eq:response-NESS-matrix,eq:response-current-matrix} by the vector $\mathbb{J}^+\boldsymbol{j}$, we arrive at
$\mathbb{R}^\pi\mathbb{J}^+\boldsymbol{j}=\boldsymbol{0}$ and $\mathbb{R}^j\mathbb{J}^+\boldsymbol{j}=\mathbb{P}\boldsymbol{j}=\mathbb{I}\boldsymbol{j}$. In coordinate form, these relations give rise to the SRRs:
\blue{
\begin{align}
\label{eq:relations}
\sum_{p\in \boldsymbol{p}}\phi_p d_p \pi_i &= 0\,,\quad
\sum_{p\in \boldsymbol{p}}\phi_p d_p j_e= j_e\,,
\end{align}
}
where $\phi_p$ are elements of the vector 
\begin{align}
\label{eq:phi}
\boldsymbol{\phi}&\equiv\mathbb{J}^+ \cdot\boldsymbol{j}\,.
\end{align}
Equations \eqref{eq:relations} generalize \cref{eq:relation-1-homog,eq:relation-2-homog} beyond homogeneous parameters \blue{with the only difference that the $h_p$'s are now replaced by the coefficients $\phi_p$.}
The structure of \cref{eq:relations} is universal, and the parameterization as well as the topology of the graph are encoded only in the vector \cref{eq:phi}, which can be \blue{easily} calculated explicitly using \cref{eq:matJ}.
\blue{The SRR for the current in \cref{eq:relations} constrains the response of the edge fluxes. It implies that all flux responses of a given edge $e$ vanish ($d_p j_e =0$ for all $p\in\boldsymbol{p}$) only when that edge is at equilibrium, i.e. when $j_e=0$.}

\begin{figure}
    \centering
    \includegraphics{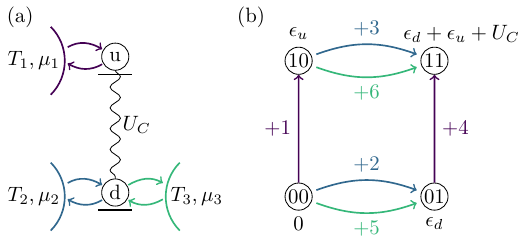}
    \caption{\blue{
    (a) Double QDs u and d coupled with three reservoirs (purple, blue, green). The reservoirs have different temperatures $T_i$ ($\beta_i=1/T_i$) and chemical potentials $\mu_i$, where $i=1,2,3$. 
    (b) Graph representation: 
    Four states $00$, $01$, $10$, and $11$ have energies $0$, $\epsilon_u$, $\epsilon_d$, and $\epsilon_d+\epsilon_u+U_C$, respectively, where $U_C$ is the Coulomb repulsion energy arising when the two dots are filled. 
    The colored arrows show the transitions assigned to the corresponding reservoir. The transition rates read $W_{\pm e}=\Gamma_{e} (1+\exp(\pm \Psi_e))^{-1}$, where $\Gamma_{e}$ are the tunneling rates and $\Psi_e$ are the edge parameters (the potential assigned to the transition $+e$): $\Psi_1=\beta_1(\epsilon_u-\mu_1)$, $\Psi_2=\beta_2(\epsilon_d-\mu_2)$, $\Psi_3=\beta_2(\epsilon_d+U_C-\mu_2)$, $\Psi_4=\beta_1(\epsilon_u+U_C-\mu_1)$, $\Psi_5=\beta_3(\epsilon_d-\mu_3)$ and $\Psi_6=\beta_3(\epsilon_d+U_C-\mu_3)$.
    }}
    \label{fig:example}
\end{figure}

\blue{\textit{Physical example:}
\cyan{
We consider the double Quantum Dots (QDs) model shown in \cref{fig:example}a~\cite{strasberg2013thermodynamics, rao2018conservation, harunari2024unveiling, sanchez2019Autonomous}. Each QD consists of a single electronic level with energy $\epsilon_u$, resp. $\epsilon_d$, that can be filled or empty due to electron exchanges with the reservoirs. Electrons cannot be transferred between the two QDs, but when the two QDs are occupied, they interact with each other via a Coulomb repulsion energy $U_C$. The four manybody states of the system and their corresponding energy are shown in \cref{fig:example}b.
}
In this case, the general explicit expressions for responses, \cref{eq:response-matrix}, hold for any parameterization. 
But to satisfy the SRRs, we must have at least $6$ independent model parameters ($N_p \geq N_e =6$ and $\mathbb{J}$ must be full row rank): The set $\{\epsilon_u, \epsilon_d, U_C\}$ is not large enough [Eq.~(B2) in \cite{noteSM}], the set $\{\epsilon_u, \epsilon_d, U_C, \mu_1, \mu_2, \mu_3\}$ is not independent as $\det\mathbb{J} = 0$ 
[Eq.~(B3) in \cite{noteSM}], but the set $\{\epsilon_u, \epsilon_d, U_C, \beta_1, \beta_2, \beta_3\}$ would work as $\det\mathbb{J} \neq 0$ [Eq.~(B4) in \cite{noteSM}].  
}

\textit{\blue{Independent} edge perturbations.---}We now restrict our theory to systems with independent parameters at every edge, namely 
\blue{
$\text{rk}~\mathbb{J}=N_p = N_e$}. 
In this case, every element $p_e$ of the vector $\boldsymbol{p}$ is assigned to its own edge $e$, which implies that the matrix $\mathbb{J}(\boldsymbol{p})=\text{diag}(\dots,\partial_{p_e}j_e,\dots)$ is diagonal and $\mathbb{J}^+=\mathbb{J}^{-1}$. Using \cref{eq:phi}, \blue{the coefficients $\phi_e$ take the explicit form} 
\begin{align}
\label{eq:phi-edge}
    \phi_e=\Big(\frac{\partial j_{e}}{\partial p_{e}}\Big)^{-1}j_e\,.
\end{align}
Since the matrix $\mathbb{J}$ is diagonal, we can rewrite \cref{eq:response-current-matrix} as 
\begin{align}
\label{eq:scaled-response}
P_{ee'}=\Big(\frac{\partial j_{e'}}{\partial p_{e'}}\Big)^{-1} \frac{d j_e}{d p_{e'}} \,.
\end{align}
This shows that the elements $P_{ee'}$ are scaled responses \blue{
where the scaling factor $\partial_{p_{e'}}j_{e'}$ is controlled by the explicit dependence $W_{\pm e}(\boldsymbol{p})$ and can be interpreted as the instantaneous response, as the edge probabilities $\pi_{s(\pm e')}$ had no time to change.
This means that $P_{ee'}$ can be seen as the ratio between the complete and instantaneous response.} 

\blue{\textit{Physical example:} For the QDs of  \cref{fig:example}, edge perturbation can be realized using the set $\boldsymbol{\Gamma}=(\dots,\Gamma_e,\dots)^\intercal$. The set $\{\epsilon_u, \epsilon_d, U_C, \beta_1, \beta_2, \beta_3\}$ also works if controlled in such a way that the edge parameters $\boldsymbol{\Psi}=(\dots,\Psi_e,\dots)^\intercal$ are changed independently, see Sec. B of \cite{noteSM}. 
}

For edge parameterization, we call the responses $d_{p_{e'}}j_e$ and $P_{ee'}$ local (resp. nonlocal) if the perturbation edge $e'$ does (resp. does not) coincide with the observation edge $e$. 
In Sec. C of \cite{noteSM}
we prove that the diagonal elements of $\mathbb{P}$ are bounded,
\begin{align}
    \label{eq:matP-diag}
    0 \leq &P_{ee}\leq 1\,,
\end{align}
which implies the bounds for the local \blue{scaled responses}\begin{align}
\label{eq:scaled-local-bound}
    \blue{0\leq\Big(\frac{\partial j_e}{\partial p_e}\Big)^{-1} \frac{d j_e}{d p_e} \leq 1}\,.
\end{align}
This result extends a previous bounds obtained in Ref.~\cite{aslyamov2024nonequilibrium} for specific parameterisations and perturbations of the rates.

Using the property of idempotent matrices $\tr~\mathbb{P}=\text{rk}~\mathbb{P}=N_{c}$, we can further derive the following CRRs:
\begin{align}
\label{eq:relation-local}
    \blue{\sum_{e=1}^{N_e} \Big(\frac{\partial j_e}{\partial p_e}\Big)^{-1} \frac{d j_e}{d p_e}= N_{c}\,.}
\end{align}
Since the l.h.s. is the sum of local scaled responses which are bounded by \cref{eq:scaled-local-bound}, \cref{eq:relation-local} show that if exactly $N_{c}$ local sensitives are saturated, then the other ones must be zero. This will be illustrated later in \cref{fig:QD}.

\begin{figure}
    \centering
    \includegraphics[width=0.5\textwidth]{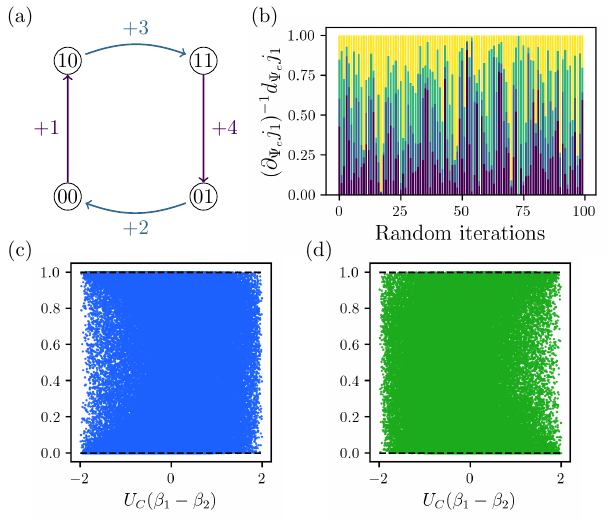}
    \caption{
    (a) Unicyclic network obtained by removing the third reservoir from \cref{fig:example}. 
    (b) The heights of the purple, blue, green, yellow bins correspond to the scaled responses of $j_1$ to perturbation of $\Psi_1,\dots,\Psi_4$ for $\Gamma_{e}$, $\beta_1$, $\beta_2$ randomly and homogeneously distributed in $0\leq \Gamma_e\leq 100$ and $0\leq \beta_1, \beta_2 \leq 2$, with $U_C=1$, $\mu_1=\mu_2=1$.  
    In (c), resp. (d), $(T_1 d_{\epsilon_u}j_1+|\partial_{\Psi_1}j_1|)/(|\partial_{\Psi_1}j_1|+|\partial_{\Psi_4}j_4|)\,$, resp.  
    $(T_2 d_{\epsilon_d}j_1+|\partial_{\Psi_3}j_3|)/(|\partial_{\Psi_2}j_2|+|\partial_{\Psi_3}j_3|)$, bounded between $0$ and $1$.
    }
    \label{fig:one-cycle}
\end{figure}

\textit{Unicyclic networks.---} We further restrict our theory to systems with a single cycle. Unicyclic systems play an important role in understanding molecular motors and metabolic networks; see e.g. the dynein model in \cite{hwang2018energetic} and the biochemical switch in \cite{owen2020universal}. 
\blue{Choosing an edge orientation such that $\boldsymbol{c}$ consists only of $1$s and $0$s and}  
using \cref{eq:matS_steady-state,eq:matP-CB-form} for the unicyclic system, we have $P_{ee'}=c_e B_{1 e'}$ and $j_e=\mathcal{J} c_e$ where $\mathcal{J}$ is the flux and $c_e$ are the elements of the single cycle $\boldsymbol{c}$, that result in $P_{ee'}=c_e/c_{e'} P_{e'e'}=j_e/j_{e'}P_{e'e'}$
Using \cref{eq:scaled-response} and
that if $j_e, j_{e'}\neq 0$ then $j_e=j_{e'}$, we find 
$P_{e'e'}=P_{ee'}=(\partial_{p_{e'}} j_{e'})^{-1}d_{p_e}j_e$
which results in the following bounds [see \cref{eq:matP-diag}]
\begin{align}
\label{eq:bounds-unicycle}
    \blue{0\leq\Big(\frac{\partial j_{e'}}{\partial p_{e'}}\Big)^{-1} \frac{d j_e}{d p_{e'}} \leq 1}
\end{align}
far any combinations of the perturbing $e'$ and observing $e$ edges. 
\blue{In addition, the SRR \eqref{eq:relations} for unicyclic network becomes
\begin{align}
\label{eq:SRR-unicycle}
    \sum_{e=1}^{N_e}\Big(\frac{\partial j_{e}}{\partial p_{e}}\Big)^{-1} \frac{d j_{e'}}{d p_{e}} = 1\,.
\end{align}
}
\blue{Since scaled responses are nonnegative and add up to one, the saturation of one automatically suppresses all the others.}

\blue{\textit{Physical example:} By removing the third reservoir (edge 5 and 6) and changing the orientation of the edges 2 and 4 in \cref{fig:example}, the model becomes a four-state unicyclic network, see \cref{fig:one-cycle}a.
The histogram in \cref{fig:one-cycle}b illustrates that the scaled responses to edge parameters $\boldsymbol{\Psi}$ are nonnegative and sum up to one as predicted by \cref{eq:bounds-unicycle,eq:SRR-unicycle}. One also sees that they are typically shared between all edges, whereas when one tends to saturate, the other ones are suppressed. 
In Sec. D of \cite{noteSM} we derive the following tight bounds for the responses of any edge current $j_e$ to the energy levels $\epsilon_u$ and $\epsilon_d$: $-|\partial_{\Psi_1}j_1|\leq T_1d_{\epsilon_u} j_e \leq |\partial_{\Psi_4}j_4|$ and $-|\partial_{\Psi_3}j_3|\leq T_2d_{\epsilon_d} j_e \leq |\partial_{\Psi_2}j_2|$, which are illustrated for different values of the thermodynamic force $U_{C}(\beta_1-\beta_2)$ in \cref{fig:one-cycle}(c,d). 
}

\textit{Multicycle systems.---}Since the elements of $\mathbb{C}$ can always be written using $\{0, 1\}$, the parametric dependence of the matrix $\mathbb{P}$ is defined by the $N_{c}N_{e}$  elements of the matrix $\mathbb{B}$ in \cref{eq:matP-CB-form}, which satisfies $\mathbb{C}\mathbb{B}\mathbb{C}=\mathbb{C}$ due to \cref{eq:matP}. Matrix $\mathbb{C}$ is full column rank and
$\mathbb{B}\mathbb{C}=\mathbb{I}_c$ is the identity matrix of size $N_c$. Defining $\overline{\mathbb{C}}$ as the invertible submatrix of $\mathbb{C}$ and noting that it is always possible to define cycles such that $\overline{\mathbb{C}}=\mathbb{I}_{c}$, 
\begin{align}
\label{eq:matB-solution-1}
 \mathbb{B}\mathbb{C}
 =\mathbb{B}
    \begin{pmatrix}
        \mathbb{I}_c \\
        \Tilde{\mathbb{C}}
    \end{pmatrix}
    =
    (\overline{\mathbb{B}},\Tilde{\mathbb{B}})
    \begin{pmatrix}
        \mathbb{I}_c \\
        \Tilde{\mathbb{C}}
    \end{pmatrix}=\mathbb{I}_{c}\,,
    \quad \overline{\mathbb{B}}=\mathbb{I}_{c}-\Tilde{\mathbb{B}}\Tilde{\mathbb{C}}\,,
\end{align}
which reduces the number of unknown elements of $\mathbb{P}$ to $\#\text{var}=N_{c}N_{e}-N_{c}^2=N_{c}(N-1)$ elements of $\Tilde{\mathbb{B}}$. 

For edge parametrisation, the fact that the scaled responses are bounded [\cref{eq:scaled-local-bound}] can be used to find the set of bounded nonlocal responses. 
For edge currents, it is equivalent to finding nondiagonal elements $P_{ee'}$ that can be written in terms of only diagonal ones $P_{ee}$ and thus be bounded. 
To do so, we define the number of independent diagonal elements as $\#\text{ide}$, which reduces the free variables of $\mathbb{P}$ to $\#\text{var}-\#\text{ide}$. 
Since there is always at least one constraint on diagonal elements because $\tr~\mathbb{P}=\text{rk}~\mathbb{P}$, we have $\#\text{ide}\leq N_{e}-1$. A greater number of constraints arise in systems with disjoint cycles (i.e. cycles that do not share edges), see Sec. E of \cite{noteSM} with an illustration for proofreading networks \cite{murugan2014discriminatory}. 
All nondiagonal elements are bounded when
\begin{align}
 \#\text{var}-\#\text{ide}=(N_{c}-1)(N-2) +(N_{e}-1-\#\text{ide}) = 0\,,    
\end{align}
which is only possible if $\#\text{ide}=N_{e}-1$, and thus if $N_{c}=1$ (unicyclic models) or if $N=2$ (two states models). 

Beyond unicyclic and two-state models, only part of the possible responses are bounded by a linear combination of the local responses. To identify which ones, using \cref{eq:matP-CB-form,eq:matB-solution-1}, we write
\begin{subequations}
\label{eq:matB-solution-2}
\begin{align}
(\mathbb{I}_c-\Tilde{\mathbb{B}}\Tilde{\mathbb{C}})_{ee}&=P_{ee}\,,\quad e\leq N_{c}\,,
\\
(\Tilde{\mathbb{C}}\Tilde{\mathbb{B}})_{ee} &= P_{ee}\,,\quad e> N_{c}\,.
\end{align}
\end{subequations}
This system of $\#\text{ide}$ equations allows us to express $\#\text{ide}$ elements $\{B_{\gamma e}^\text{lin}\}$ as a linear combination of the bounded diagonal elements $\{P_{e e}\}$.
The other elements $\{B_{\gamma e}\}\setminus\{B_{\gamma e}^\text{lin}\}$ are not restricted by the bounds in \cref{eq:matP-diag}. 
Thus, the elements $P_{ee'}=\sum_{\gamma}C_{e\gamma}B_{\gamma e'}$ are therefore bounded if they contain only terms from the set $\{B_{\gamma e}^\text{lin}\}$. This will be illustrated in \cref{fig:QD}. 

\textit{Thermodynamic responses.---}The rates can be expressed in terms of their symmetric $v_{e} \equiv \ln\sqrt{W_{+e} W_{-e}}=v_{-e}$ and antisymmetric $w_{e} \equiv \ln\sqrt{W_{+ e}/W_{- e}}=-w_{-e}$ parts as $W_{\pm e}=\exp(v_e\pm w_e)$. \blue{For the QDs of \cref{fig:example}, $v_e=\ln \Gamma_e$ and $w_{e}=-\ln(1+\exp(\Psi_e))$.}
This decomposition is used in stochastic thermodynamics (i.e. for open systems undergoing transitions caused by thermal reservoirs) because the antisymmetric part defines the energetics of the system~\cite{rao2018conservation,falasco2023macroscopic}:
\begin{align}
\label{eq:LDB-general}
   \ln\frac{W_{+e}}{W_{-e}} = \sum_{i\in\mathcal{S}} E_i S_{ie} + \mathcal{F}_e \,,\quad \mathcal{F}_e &= \sum_{\alpha \in \mathcal{R}} X_{e\alpha} f_{\alpha}\,.
\end{align}
The l.h.s. of \cref{eq:LDB-general} is the entropy change in the reservoirs caused by a transition $e$. It can always be split, in the r.h.s., as a change in the Massieu potential $E_i$ of the state $i$ and as a non-conservative contribution $\mathcal{F}_{e}=-\mathcal{F}_{-e}$. 
The latter can be expressed, in \cref{eq:LDB-general}, as a sum over a subset of reservoirs $\mathcal{R}$ where each term consists
of an amount of conserved quantities exchanged with the reservoirs during a transition $e$, $X_{e\alpha}$, multiplied by a conjugated (fundamental \cite{rao2018conservation}) thermodynamic force, $f_\alpha$ made of differences of intensive fields of the reservoirs such as inverse temperatures or chemical potentials \cite{rao2018conservation}.
Using \cref{eq:LDB-general} and \cref{eq:scaled-response}, we find 
\begin{align}
\label{eq:response-fundamental}
d_{f_\alpha}j_e&=\sum_{\rho\in\mathcal{E}}d_{f_\alpha} \mathcal{F}_{\rho} d_{\mathcal{F}_\rho}j_{e}
=\sum_{\rho\in\mathcal{E}}X_{\rho\alpha}P_{e\rho}\partial_{\mathcal{F}_\rho}j_{\rho}\;,
\end{align}
which can be obtained analytically for known $\partial_{\Psi_\rho}j_{\rho}$. Since the (fundamental \cite{rao2018conservation}) currents exchanged with the different reservoirs are the elements of the vector $\boldsymbol{I}=\mathbb{X}^\intercal j$, their response to the thermodynamic forces read
\begin{align}
\label{eq:response-fundamental-matrix}
    \mathbb{R}^I\equiv[d_{f_{\alpha'}}I_{\alpha}]_{\{\alpha,\alpha'\}}=\mathbb{X}^\intercal\mathbb{P}\mathbb{J}(\boldsymbol{\mathcal{F}})\mathbb{X}=\mathbb{X}^\intercal\mathbb{R}^j\mathbb{X}\,,
\end{align}
where $\alpha,\alpha'\in \mathcal{R}$ and $\mathbb{J}(\boldsymbol{\mathcal{F}})=\text{diag}(\dots,\partial_{\mathcal{F}_e}j_{e},\dots)$. 
\blue{
Close to equilibrium, $\mathbb{R}^I$ reduces to the semi-positive definite Onsager matrix (Sec. F of \cite{noteSM}).
The lack of symmetry of $\mathbb{R}^I$ can thus be measured experimentally as $|R^{I}_{\alpha\alpha'}-R^{I}_{\alpha'\alpha}|$ and used to determine if the system is far from equilibrium. 
}
\\
Let us now assume that $w_{e}$ is independent of $v_e$. This is relevant, for example, for Arrhenius-like rates \cite{aslyamov2024nonequilibrium}, as well as for electron transfer rates in CMOS transistors \cite{gopal2022Large} or single electron tunneling rates in the wide-band approximation \cite{sanchez2019Autonomous}.
\Cref{eq:LDB-general} shows that $w_e$ depends only on the perturbation of the energy and forces, but does not depend on the perturbation of the kinetic parameters.
Such kinetic $v_e$ and energy (thermodynamic forces) $w_e$ perturbations will be constrained by \cref{eq:relations}. 
Calculating the partial derivatives $\partial_{w_e}j_e=\tau_e$ and $\partial_{v_e}j_e=j_e$, we find $\phi_e=j_e/\tau_e$ (resp. $\phi_e=1$) for $w_e$ (resp. $v_e$), where $\tau_e \equiv W_e\pi_{s(+e)}+W_{-e}\pi_{s(-e)}$ is the edge traffic.
Inserting $\phi_e$ into \cref{eq:relations}, we arrive at the Symmetric and Antisymmetric SRRs 
\begin{subequations}
\label{eq:relations-sym-antisym}
\begin{align}
    \sum_{e}\frac{j_e}{\tau_e}d_{w_e}\boldsymbol{\pi}&=\boldsymbol{0}\,,
    \quad \sum_{e}d_{v_e}\boldsymbol{\pi}=\boldsymbol{0}\,,\\
    \label{eq:relation-2-arrhenius}
    \sum_{e}\frac{j_e}{\tau_e}d_{w_e}\ln\boldsymbol{j}&=\boldsymbol{1}\,, \quad \sum_{e}d_{v_e}\ln\boldsymbol{j}=\boldsymbol{1}\,.
\end{align}
\end{subequations}
We note that unlike the antisymmetric parameterisation, the symmetric one is homogeneous $\boldsymbol{h}=(\dots,v_e,\dots)^\intercal$ as the symmetric RSSs in \cref{eq:relations-sym-antisym} coincide with \cref{eq:relation-1-homog,eq:relation-2-homog}. 

\begin{figure}
    \centering
    \includegraphics[width=\columnwidth]{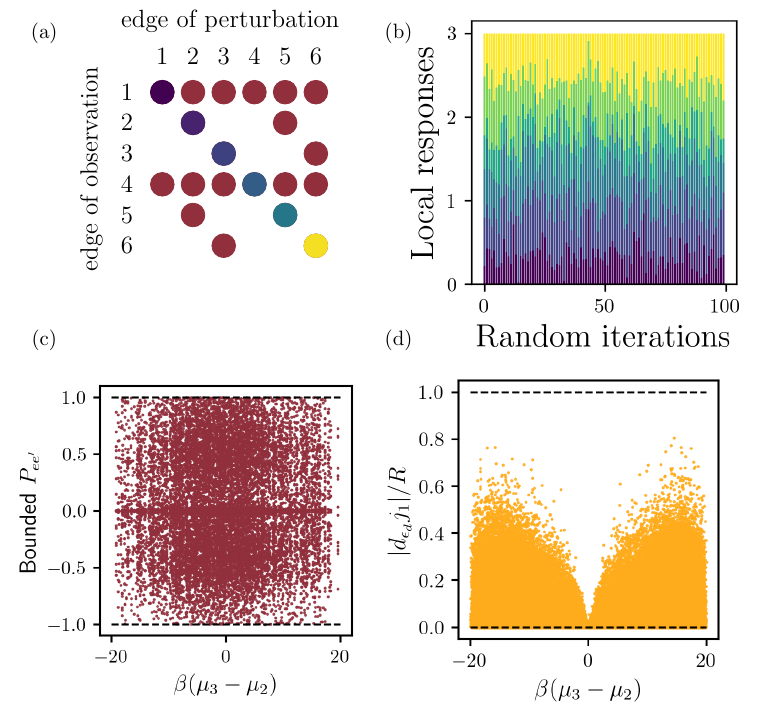}
    \caption{
    \blue{For the model in \cref{fig:example}: (a) The disks indicate the $20$ bounded $P_{ee'}$'s out of $36$. 
    (b) Validity of the CRR [\cref{eq:relation-local}] with $N_c=3$. The heights of the color bins (from black to yellow) correspond to $P_{ee}$ for $e=1,\dots,6$ and to
    randomly and homogeneously distributed $0 \leq \epsilon_u, \epsilon_d, U_C\leq 5$, $0<\Gamma_e\leq 1000$ and $-10 \leq \mu_2,\mu_3 \leq 10$, with $T_i=1$, $\mu_1=0$.
    (c) $P_{ee'}$ corresponding to the red disks in $(a)$, as a function of the thermodynamic force $\beta(\mu_3-\mu_2)$.
    (d) Physical responses $|d_{\epsilon_d}j_1|$ in units of $R=\sum_{e\neq 1,4}|\partial_{\Psi_e}j_e|$. Dashed lines denote our bounds.}
    }
    \label{fig:QD}
\end{figure}

\blue{\textit{Physical example:}
For the QDs in \cref{fig:example}, we use \cref{eq:matB-solution-2} to find all the elements $P_{ee'}$ which are linear combinations of diagonal elements and are thus bounded, see Sec. G of \cite{noteSM} and \cref{fig:QD}a. The $6$ local scaled responses, $P_{ee}$, sum to $N_c=3$ as predicted by the CRR [\cref{eq:relation-local}], see \cref{fig:QD}b. The nonlocal scaled responses can be negative, but those marked as disks in \cref{fig:QD}a are bounded as $-1\leq P_{ee'}\leq 1$, see Eq.~(G4) in \cite{noteSM} and \cref{fig:QD}c. 
}
\blue{We use the properties of the matrix $\mathbb{P}$ to bound the responses of the current to physical parameters in Sec. H of \cite{noteSM}.
We find that $d_p j_{\alpha}=\sum_{e}P_{\alpha e}\partial_p \Psi_e$ for $\alpha={1,4}$, where $-1 \leq P_{\alpha e}\leq 1$. We also find $|d_{\epsilon_d}j_\alpha|\leq R$, where $R=|\partial_{\Psi_2}j_2|+|\partial_{\Psi_3}j_3|+|\partial_{\Psi_5}j_5|+|\partial_{\Psi_6}j_6|$. This is illustrated numerically in \cref{fig:QD}d for different values of the thermodynamic force $\beta (\mu_3-\mu_2)$, where we see that large responses arise far from equilibrium.  
}

\textit{Future studies.---} Our approach provides powerful tools to identify networks that are highly sensitive or extremely resilient to perturbations. It is also ideally suited to study the responses of enzymatic changes (proofreading, sensing) in chemical reaction networks, in particular in conjunction with recently developed circuit theory \cite{avanzini2023circuit}. 
\blue{Extending our approach to non-stationary response theory of Markov processes as in Refs.~\cite{mitrophanov2003stability,mitrophanov2005sensitivity,harunari2024mutual,zheng2024information} is also an interesting perspective. 
}
\\
\textit{Acknowledgements---}This research was supported by Fonds National de la Recherche—FNR, Luxembourg: M. E. and T.A. by Project ChemComplex (C21/MS/16356329) and T. A. also by Project ThermoElectroChem (C23/MS/18060819).

\newpage
\appendix

\section{Derivation of Eq.~6.}
\label{sec:appendix-response}
Here we derive the matrix expressions (6) for the probability and current responses to an arbitrary parameterisation $\boldsymbol{p}$. 
The total derivative $d_{p}$ of Eq.~(2) on $p\in\boldsymbol{p}$ reads
\begin{align}
    \label{eq:deta-1}
    \mathbb{S}(\partial_{p}\mathbb{\Gamma})\boldsymbol{\pi}+\mathbb{S}\mathbb{\Gamma}d_{p}\boldsymbol{\pi}=\boldsymbol{0}\,.
\end{align}
Using $\sum_{i=1}^N\pi_i=1$ we find one of the derivatives, say the $N$th one, as $d_{p}\pi_N=-\sum_{i=1}^{N-1}d_p\pi_i$. Inserting it in the coordinate form of \cref{eq:deta-1} we arrive at:
\begin{align}
\footnotesize
    \label{eq:deta-2}
    \sum_{j\in\Hat{\mathcal{S}}} \sum_{e\in\mathcal{E}}S_{ie}(\Gamma_{ej}-\Gamma_{eN})d_{p}\pi_j&=-\sum_{e\in\mathcal{E}}\sum_{k\in\mathcal{S}}S_{ie}(\partial_{p}\Gamma_{ek})\pi_k\,, 
\end{align}
where $\Hat{\mathcal{S}}=\mathcal{S}\setminus\{N\}$ denotes the space of $N-1$ states, or 
\begin{align}
\footnotesize
    \label{eq:deta-2bis}
    \sum_{j\in\Hat{\mathcal{S}}}K_{ij}d_{p}\pi_j=-\sum_{e\in\mathcal{E}}S_{ie}J_{ep}\,,
\end{align}
where $J_{ep}=\partial_p j_e$ are elements of the Jacobian defined in Eq.~(7) and where we introduced the matrix
\begin{align}
\footnotesize
\label{eq:matK}
   \mathbb{K}
   &=\Big[\sum_{e\in\mathcal{E}}S_{ie}(\Gamma_{ej}-\Gamma_{eN})\Big]_{i,j\in\Hat{\mathcal{S}}}=\Hat{\mathbb{S}}\Hat{\mathbb{\Gamma}}\,,
\end{align}
with $\Hat{\mathbb{S}}=[S_{ie}]_{i\in\Hat{\mathcal{S}},e\in\mathcal{E}}$ and  $\Hat{\mathbb{\Gamma}}=[\Gamma_{ej}-\Gamma_{eN}]_{e\in\mathcal{E},j\in\Hat{\mathcal{S}}}$. 

We proceed with the proof that the matrix $\mathbb{K}$ is invertible.
The auxiliary matrix $\mathbb{K}_N$ coincides with $\mathbb{W}$, except the $N$th row, which is the row of ones. 
Therefore $\det\mathbb{K}_N=$
\begin{align}
\footnotesize
\det \begin{pmatrix}
    [W_{m,n}]_{m\in\Tilde{S},n\in\mathcal{S}} \\
    \boldsymbol{1}^\intercal
    \end{pmatrix}  
=\det \begin{pmatrix}
    \mathbb{K} & [W_{mN}]_{m\in\Tilde{S}}\\
    \boldsymbol{0}^\intercal & 1
    \end{pmatrix}
=\det\mathbb{K}\,.
\end{align}
Then, using the results of Ref.~\cite{aslyamov2024nonequilibrium} for $\det\mathbb{K}_N$, we arrive at:
\begin{align}
\label{eq:matK-det}
    \det\mathbb{K}=\det\mathbb{K}_N=\prod_{i=1}^{N-1}\lambda_i\neq0\,,
\end{align}
where $\lambda_i$ are nonzero eigenvalues of the matrix $\mathbb{W}$, which implies that $\mathbb{K}$ is invertible.  

Multiplying \cref{eq:deta-2bis} by $\mathbb{K}^{-1}$, we find the responses $d_p\pi_i$. Using them to calculate $d_p\pi_N$ and rewriting the result in matrix form, we arrive at Eq.~(6a). 
Similarly, using \cref{eq:deta-1} we can calculate the response of the current:
\begin{align}
\label{eq:response-current}
d_{p}\boldsymbol{j}&=\partial_{p}\boldsymbol{j}+\boldsymbol{\Gamma}\partial_{p}\boldsymbol{\pi}=\partial_{p}\boldsymbol{j}+\Big[\sum_{i\in\hat{\mathcal{S}}}(\Gamma_{ei}-\Gamma_{eN})d_{p}\pi_i\Big]_{e\in\mathcal{E}}\nonumber\\
&=(\mathbb{I}-\Hat{\mathbb{\Gamma}}(\Hat{\mathbb{S}}\Hat{\mathbb{\Gamma}})^{-1}\Hat{\mathbb{S}})\partial_{p}\boldsymbol{j}=\mathbb{P}\partial_{p}\boldsymbol{j}\,,
\end{align}
which has the matrix form shown in Eq.~(6b). 

\section{Physical Parameters in Fig.~1}
To calculate the Jacobian (7) for physical parameter $p$ from the model of Fig.~1, we apply the chain rule:
\begin{align}
    \mathbb{J}&=\Big[\partial_p j_e\Big]_{\{e,p\}} = \Big[\sum_{e'}\partial_{\Psi_{e'}} j_e\partial_p \Psi_{e'}\Big]_{\{e,p\}}=
    \Big[\partial_{\Psi_{e}} j_e\partial_p \Psi_{e}\Big]_{\{e,p\}}\,,
\end{align}
where we used that $j_{e}$ explicitly depends on $\Psi_e$ only. 

(i) The set $\boldsymbol{p}_1=\{\epsilon_u,\epsilon_d,U_c\}$ results in the Jacobian
\begin{align}
    \mathbb{J}_1 =
    \begin{pmatrix}
        \beta_1\partial_{\Psi_1}j_1 & 0 & 0 \\
        0 & \beta_2 \partial_{\Psi_2}j_2 & 0\\
        0 & \beta_2 \partial_{\Psi_3}j_3 & \beta_2\partial_{\Psi_3}j_3 \\
       \beta_1 \partial_{\Psi_4}j_4 & 0 &  \beta_1\partial_{\Psi_4}j_4\\
        0 &  \beta_3\partial_{\Psi_5}j_5 & 0\\
        0 &  \beta_3\partial_{\Psi_6}j_6 & \beta_3\partial_{\Psi_6}j_6
    \end{pmatrix}\,,
\end{align}
which has the rank $\text{rk}~\mathbb{J}_1=3$. Since $\text{rk}~\mathbb{J}_1<N_e=6$, the set $\boldsymbol{p}_1$ does not satisfy the conditions of the response relations [Eq.~10].

(ii) The Jacobing of the set $\boldsymbol{p}_2=\{\epsilon_u,\epsilon_d,U_c,\mu_1, \mu_2, \mu_3\}$ reads
\begin{align}
    \mathbb{J}_2 =
    \begin{pmatrix}
        \beta_1\partial_{\Psi_1}j_1 & 0 & 0 & - \beta_1\partial_{\Psi_1}j_1 & 0 & 0 \\
        0 & \beta_2 \partial_{\Psi_2}j_2 & 0 &  0 & -\beta_2 \partial_{\Psi_2}j_2 & 0 \\
        0 & \beta_2 \partial_{\Psi_3}j_3 & \beta_2\partial_{\Psi_3}j_3 & 0 & -\beta_2 \partial_{\Psi_3}j_3 & 0 \\
       \beta_1 \partial_{\Psi_4}j_4 & 0 &  \beta_1\partial_{\Psi_4}j_4 & -\beta_1 \partial_{\Psi_4}j_4 & 0 & 0\\
        0 &  \beta_3\partial_{\Psi_5}j_5 & 0 & 0 &  0 & -\beta_3\partial_{\Psi_5}j_5\\
        0 &  \beta_3\partial_{\Psi_6}j_6 & \beta_3\partial_{\Psi_6}j_6 &  0 &  0 & - \beta_3\partial_{\Psi_6}j_6
    \end{pmatrix}\,,
\end{align}
where $N_p=N_e$, but $\text{rk}~\mathbb{J}_2=4<N_e=6$ means that $\boldsymbol{p}_2$ does not work for the response relations.

(iii) For the set $\boldsymbol{p}_2=\{\epsilon_u,\epsilon_d,U_c,\beta_1, \beta_2, \beta_3\}$ we have
\begin{widetext}
\begin{align}
    \mathbb{J}_3 =
    \begin{pmatrix}
        \beta_1\partial_{\Psi_1}j_1 & 0 & 0 & (\epsilon_u-\mu_1)\partial_{\Psi_1}j_1 & 0 & 0 \\
        0 & \beta_2 \partial_{\Psi_2}j_2 & 0 &  0 & (\epsilon_d-\mu_2) \partial_{\Psi_2}j_2 & 0 \\
        0 & \beta_2 \partial_{\Psi_3}j_3 & \beta_2\partial_{\Psi_3}j_3 & 0 & (\epsilon_d+U_C-\mu_2)\partial_{\Psi_3}j_3 & 0 \\
       \beta_1 \partial_{\Psi_4}j_4 & 0 &  \beta_1\partial_{\Psi_4}j_4 & (\epsilon_u+U_C-\mu_1) \partial_{\Psi_4}j_4 & 0 & 0\\
        0 &  \beta_3\partial_{\Psi_5}j_5 & 0 & 0 &  0 & -(\epsilon_d-\mu_3)\partial_{\Psi_5}j_5\\
        0 &  \beta_3\partial_{\Psi_6}j_6 & \beta_3\partial_{\Psi_6}j_6 &  0 &  0 & (\epsilon_d+U_C-\mu_3)\partial_{\Psi_6}j_6
    \end{pmatrix}\,,
\end{align} 
\end{widetext}
which satisfies the conditions of summation relations with $\text{rk}~\mathbb{J}_3=6=N_e$. In addition for $\boldsymbol{p}_2$ the following matrix is nonsingular:
\begin{align}
    \det \big([\partial_{p_{e'}}\Psi_e]_{\{e,e'\}}\big) \neq 0\,,
\end{align}
which implies that controlling $d\boldsymbol{p}$ one can change $d\boldsymbol{\Psi}$ independently at every edge. Thus, for Fig.~1 it is possible to use $\boldsymbol{\Psi}$ as an edge perturbation. Since $\mathbb{J}(\boldsymbol{\Psi})$ and $\mathbb{J}(\boldsymbol{\Gamma})$ are diagonal, the two scaled responses are identical $P_{ee'} = (\partial_{\Gamma_{e'}}j_{e'})^{-1} d_{\Gamma_{e'}}j_e=(\partial_{\Psi_{e'}}j_{e'})^{-1}d_{\Psi_{e'}}j_e$.

\section{Proof of Eq.~(15)}
Here we prove the bounds in Eq.~(15) of the main text. We introduce the projection matrix $\mathbb{\Pi}=\mathbb{I}-\mathbb{P}$ with the diagonal elements defined by
\begin{align}
\label{eq:matp}
   \Pi_{ee}=\sum_{x,x'\in\Hat{\mathcal{S}}}\Hat{\Gamma}_{ex}(\mathbb{K}^{-1})_{xx'}S_{x'e}\,.
\end{align}
One can see that Eq.~(15) is equivalent to $0 \leq \mathbb{\Pi}_{ee}\leq 1$. Our proof strategy is the following: first, we prove that all diagonal elements $\Pi_{ee}$ are positive; then we show the upper bound $\Pi_{ee}\leq 1$. 
To do so for all edges $e$ we distinguish two possible cases: (i) the edge $e$ contains the state $N$ as $s(+e)=N$ or $s(-e)=N$; (ii) the edge $s(+e)=n$ and $s(-e)=m$, where $n,m\neq N$.

\textit{(i) Edges with $s(+e)=N$ or $s(-e)=N$.}---Let us consider the edge $e$ with $s(+e)=N$ and $s(-e)=n$ which corresponds $S_{Ne}=-1$, $\Gamma_{eN}=W_{+e}$ and $S_{ne}=1$, $\Gamma_{e n}=-W_{-e}$. 
We note that all elements of the row $[\Gamma_{ex}-\Gamma_{eN}]_{x\in \Hat{\mathcal{S}}}$ are nonzero due to $\Gamma_{eN}\neq 0$.
Then \cref{eq:matp} reads
\begin{align}
\label{eq:matp-diag-case-1}
    \Pi_{ee}=&\sum_{x\in\Hat{\mathcal{S}}}\Hat{\Gamma}_{ex}(\mathbb{K}^{-1})_{xn}S_{ne}\nonumber\\
    =&-W_{-e}(\mathbb{K}^{-1})_{nn}-W_{+e}\sum_{x\in\Hat{\mathcal{S}}}(\mathbb{K}^{-1})_{xn}\nonumber\\
    =&
    \underbrace{-W_{-e}\frac{M_{nn}(\mathbb{K})}{\det\mathbb{K}}}_{T_1}
    \underbrace{-W_{+e}\sum_{x\in\Hat{\mathcal{S}}}\frac{(-1)^{n+x}M_{nx}(\mathbb{K})}{\det\mathbb{K}}}_{T_2}
    \,,
\end{align}
where the $M_{ij}(\mathbb{K})$ is the $ij$th minor of the matrix $\mathbb{K}$ and where we use $(\mathbb{K}^{-1})_{ij}=(-1)^{i+j}M_{ji}(\mathbb{K})/\det{\mathbb{K}}$. Using Eq.~(A3) we write the elements of $\mathbb{K}$ as the following:
\begin{align}
\label{eq:matK-elements}
    K_{ij}=\sum_{e'\in\mathcal{E}}S_{ie'}(\Gamma_{e'j}-\Gamma_{e'N})\,,
\end{align}
and proceed analyzing the dependence of the elements in \cref{eq:matK-elements} on the rates $W_{\pm e}$. Using \cref{eq:matK-elements} for $i,j\in\hat{\mathcal{S}}$ we notice that only one element of $\mathbb{K}$ depends on $\Gamma_{en}=-W_{-e}$ namely
\begin{align}
\label{eq:matK-nn}
    K_{nn}&=S_{ne}(\Gamma_{en}-\Gamma_{e N})+\dots=-W_{-e}+\dots\,,
\end{align}
where the symbol $\dots$ denotes the contribution which does not depend on the parameter of interest (in this case $W_{-e}$). Also, \cref{eq:matK-elements} shows that $\Gamma_{eN}=W_{+e}$ contributes to all elements $K_{nj}$ in the $n$th row of the matrix $\mathbb{K}$: 
\begin{align}
\label{eq:matK-nj}
   K_{nj}&=-W_{+e}+\dots\,,
\end{align}
where $\dots$ denotes the terms which do not depend on $W_{+e}$. Calculating the determinant of $\mathbb{K}$ with respect to the $n$th row, we find
\begin{align}
\label{eq:det-matK-nj}
\footnotesize
\det{\mathbb{K}}&=\sum_{j}(-1)^{n+j}K_{nj}M_{nj}(\mathbb{K})\\
&=-W_{-e}M_{nn}(\mathbb{K})-W_{+e}\sum_{x\in\Hat{\mathcal{S}}}(-1)^{n+x}M_{nx}(\mathbb{K})+Q_1\,,\nonumber
\end{align}
where $Q_1$ does not depend on $W_{\pm e}$.
Furthermore, since the $n$th row of $\mathbb{K}$ does not contribute to the minors $M_{nj}(\mathbb{K})$, they do not depend on $W_{\pm e}$.

The sign of the determinant is defined by the size of the system as $\sgn\det{\mathbb{K}}=(-1)^{N-1}$; see Eq.~(A5). Since $Q_1$ does not depend on $W_{\pm e}$, calculating the sign of \cref{eq:det-matK-nj} with $W_{\pm e}=0$ we arrive at
\begin{align}
    \label{eq:Q1-sign}
    \sgn~Q_1 = \sgn\det\mathbb{K} = (-1)^{N-1}\,. 
\end{align}

We notice that the sign of the term $T_1$ does not depend on $W_{-e}$. We use this property to find the sign of $T_1$ considering the limit $W_{-e}\to \infty$:
\begin{align}
    \label{eq:sign-T1}
    \sgn~T_1=-\sgn\lim_{W_{-e}\to\infty}W_{-e}\frac{M_{nn}(\mathbb{K})}{\det\mathbb{K}}=1\,,
\end{align}
where we use $\det\mathbb{K}\to-W_{-e}M_{nn}$ as the rate $W_{-e}$ tends to infinity [see \cref{eq:det-matK-nj}]. Similarly, we find $\sgn~T_2$ calculating the limit $W_{+e}\to \infty$:
\begin{align}
    \label{eq:sign-T2}
    \sgn~T_2=-\sgn\lim_{W_{+e}\to\infty}W_{+e}\sum_{x\in\Hat{\mathcal{S}}}\frac{(-1)^{n+x}M_{nx}(\mathbb{K})}{\det\mathbb{K}}=1\,.
\end{align}
Combining \cref{eq:sign-T1,eq:sign-T2,eq:matp-diag-case-1} we arrive at $\Pi_{ee}\geq 0$ for $s(+e)=N$. The case $s(-e)=N$ can be considered in a similar way. Thus, we have $\Pi_{ee}\geq 0$ for the edges of case (i). To find an upper bound, we rewrite \cref{eq:matp-diag-case-1} using \cref{eq:det-matK-nj,eq:Q1-sign} as:
\begin{align}
\label{eq:bound-case-1}
    \Pi_{ee}=1-\Bigg|\frac{Q_1}{\det\mathbb{K}}\Bigg|\leq 1
    \,.
\end{align}
Thus, $\Pi_{ee}\geq 0$ and \cref{eq:bound-case-1} result in $0 \leq \Pi_{ee}\leq 1$ for case (i). In terms of the original matrix $\mathbb{P}$ we have  $0 \leq P_{ee}\leq 1$.

\textit{(ii) Edges with $s(\pm e)\neq N$.}---Let us consider the edge $e$ with $s(+e)=n$ and $s(-e)=m$, where $n,m\neq N$. In this case, we have $\Gamma_{eN}=0$ and the row $[\Gamma_{ex}-\Gamma_{eN}]_{x\in\Hat{\mathcal{S}}}$ contains only two non-zero elements $\Gamma_{en}=W_{+e}$ and $\Gamma_{em}=-W_{-e}$; and the column $[S_{xe}]_{x\in\Hat{\mathcal{S}}}$ has two nonzero elements $S_{ne}=-1$ and $S_{me}=1$.
Therefore, the diagonal elements read
\begin{align}
\footnotesize
\label{eq:matp-diag-case-2}
    \Pi_{ee}=&\sum_{i,j=n,m}\Gamma_{e i}(\mathbb{K}^{-1})_{ij}S_{j e}\nonumber\\
    =&\underbrace{-W_{+e}\frac{M_{nn}(\mathbb{K})}{\det\mathbb{K}}
    +(-1)^{n+m}W_{+e}\frac{M_{mn}(\mathbb{K})}{\det\mathbb{K}}}_{T_3}\nonumber\\
    &\underbrace{+(-1)^{m+n}W_{-e}\frac{M_{nm}(\mathbb{K})}{\det\mathbb{K}}-W_{-e}\frac{M_{mm}(\mathbb{K})}{\det\mathbb{K}}}_{T_4}
    \,.
\end{align}
To proceed, we analyze the dependence of the matrix $\mathbb{K}$ on the rates $W_{\pm e}$. We find that two elements of  $\mathbb{K}$ depend on $W_{+e}$ namely
\begin{subequations}
\label{eq:matK-W+e}
\begin{align}
    K_{nn}&=S_{ne}\Gamma_{en}+\dots=-W_{+e}+\dots\,,\\
    K_{mn}&=S_{me}\Gamma_{en}+\dots=W_{+e}+\dots\,,
\end{align}
\end{subequations}
and another two elements depend on $W_{-e}$ namely
\begin{subequations}
\label{eq:matK-W-e}
\begin{align}
    K_{nm}&=S_{ne}\Gamma_{em}+\dots=W_{-e}+\dots\,,\\
    K_{mm}&=S_{me}\Gamma_{em}+\dots=-W_{-e}+\dots\,.
\end{align}
\end{subequations}
Therefore, the dependence of the matrix $\mathbb{K}$ on $W_{+e}$ (resp. $W_{-e}$) is encoded in the $n$th column (resp. $m$th column). For this reason, the minors $M_{nn}(\mathbb{K})$ and $M_{mn}(\mathbb{K})$ (resp. $M_{nm}(\mathbb{K})$ and $M_{mm}(\mathbb{K})$) do not depend on $W_{+e}$ (resp. $W_{-e}$). Thus, the signs $\sgn~T_3$ and $\sgn~T_4$ can be found by considering the limits $W_{+e}\to\infty$ and $W_{-e}\to\infty$, respectively. 

For $T_3$ we calculate the $\det{\mathbb{K}}$ in the respect to the $n$th column and use \cref{eq:matK-W+e}:
\begin{align}
\label{eq:detK-W+e}
    \det{\mathbb{K}}&=\sum_{i}(-1)^{i+n}K_{in}M_{in}(\mathbb{K})\nonumber\\
    &=-W_{+e}M_{nn}(\mathbb{K})+(-1)^{n+m}W_{+e}M_{mn}(\mathbb{K})+\dots\,,
 \end{align}
where the dots denote the contribution which does not depend on $W_{+e}$.
Inserting \cref{eq:detK-W+e} in $T_3$ from \cref{eq:matp-diag-case-2} and calculating the sign in the limit $W_{+e}\to\infty$ we arrive at
\begin{align}
\label{eq:sign-T3}
    \sgn~T_3=\lim_{W_{+e}\to \blue{\infty}}\frac{-W_{+e}M_{nn}(\mathbb{K})+(-1)^{n+m}W_{+e}M_{mn}(\mathbb{K})}{\det\mathbb{K}}=1\,.
\end{align}

For $T_4$ we calculate $\det{\mathbb{K}}$ in the respect to the $m$th column and use \cref{eq:matK-W-e} to find
\begin{align}
\label{eq:detK-W-e}
    \det{\mathbb{K}}&=\sum_{i}(-1)^{i+n}K_{in}M_{im}(\mathbb{K})\nonumber\\
    &=(-1)^{n+m}M_{nm}W_{-e}(\mathbb{K})-W_{-e}M_{mm}(\mathbb{K})+\dots\,,
 \end{align}
 where the dots denote the terms which do not depend on $W_{-e}$. 
Inserting \cref{eq:detK-W-e} in $T_4$ from \cref{eq:matp-diag-case-2} and calculating the sign in the limit $W_{-e}\to\infty$ we arrive at
\begin{align}
\label{eq:sign-T4}
    \sgn~T_4=\lim_{W_{-e}\to \blue{\infty}}\frac{(-1)^{n+m}W_{-e}M_{nm}(\mathbb{K})-W_{-e}M_{mm}(\mathbb{K})}{\det\mathbb{K}}=1\,.
\end{align}
Using \cref{eq:sign-T3,eq:sign-T4,eq:matp-diag-case-2}, we prove that all diagonal elements $\Pi_{ee}\geq 0$ are positive. 

Finally, we find an upper bound for $\Pi_{ee}$.
We modify the matrix $\mathbb{K}$ adding the $m$th row to the $n$th row. In the result, the new $n$th row does not depend on $W_{\pm e}$; see \cref{eq:matK-W+e,eq:matK-W-e}. This operation does not change the determinant $\det\mathbb{K}$ that allows to write it using the $m$th row
\begin{align}
\label{eq:det-matK-linear}
\det\mathbb{K} = W_{+e} \alpha +  W_{-e} \beta + Q_2 \,,
\end{align}
where $\alpha$, $\beta$ and  $Q_2$ are quantities which do not depend on $W_{\pm e}$. \Cref{eq:det-matK-linear} shows that $\det\mathbb{K} $ is the linear function of $W_{\pm e}$
From the other hand, an explicit dependence of $\det\mathbb{K}$ on $W_{+e}$ (resp. on $W_{-e}$) is described by \cref{eq:detK-W+e} (resp. \cref{eq:detK-W-e}). 
Thus \cref{eq:det-matK-linear} must has the following form
\begin{align}
\label{eq:det-matK-linear-2}
\det\mathbb{K} =& W_{+e}(-M_{nn}(\mathbb{K})+(-1)^{n+m}M_{mn}(\mathbb{K})) +\nonumber\\  
&+W_{-e}((-1)^{n+m}M_{nm}(\mathbb{K})-M_{mm}(\mathbb{K})) + Q_2 \,.
\end{align}
Since $Q_2$ does not depend on $W_{\pm e}$ we have  $\sgn\det\mathbb{K}=\sgn~Q_2$, which gives us:
\begin{align}
\label{eq:bound-case-2}
    \Pi_{ee}=1-\Bigg|\frac{Q_2}{\det\mathbb{K}}\Bigg|\leq 1
    \,,
\end{align}
where we used \cref{eq:matp-diag-case-2,eq:det-matK-linear-2}. Combining $\Pi_{ee}\geq 0$ and \cref{eq:bound-case-2} we have $0 \leq \Pi_{ee}\leq 1$ for case (ii).  
Moving back to the original projection matrix $\mathbb{P}=\mathbb{I}-\mathbb{\Pi}$, we prove Eq.~(15).

\section{Physical Perturbations: Unicyclic}
Here we show an example of how the properties of matrix $\mathbb{P}$ can be used to bound the current responses in a unicyclic network to physical parameters. We consider the network from Fig.~2 with 
$\Psi_1=\beta_1(\epsilon_u-\mu_1)$, 
$\Psi_2=-\beta_2(\epsilon_d-\mu_2)$, 
$\Psi_3=\beta_2(\epsilon_d+U_C-\mu_2)$, 
$\Psi_4=-\beta_1(\epsilon_u+U_C-\mu_1)$ 
and $W_{\pm e}=\Gamma_e(1+\exp(\pm \Psi_e))^{-1}$, where the signs $\Psi_2$ and $\Psi_4$ are different from those in Fig.~1 due to the orientation changes. For example, we calculate the responses $d_{\epsilon_u}j_e$ and $d_{\epsilon_d}j_e$. From Eq.~(6b) we have 
\begin{align}
\label{eq:appendix-unicyclic-1}
    \begin{pmatrix}
        d_{\epsilon_u}j_e & d_{\epsilon_d}j_e
    \end{pmatrix}
    =
    \begin{pmatrix}
        P_{11} & P_{22} & P_{33} & P_{44}
    \end{pmatrix}
    \begin{pmatrix}
        \beta_1\partial_{\Psi_1}j_1 & 0 \\
        0 & -\beta_2\partial_{\Psi_2}j_2 \\
        0 & \beta_2\partial_{\Psi_3}j_3 \\
        -\beta_1\partial_{\Psi_4}j_4 & 0
    \end{pmatrix}\,,
\end{align}
where we used the property $P_{ee'}=P_{e'e'}$ for unicyclic networks. In terms of the elements, \cref{eq:appendix-unicyclic-1} reads
\begin{subequations}
\begin{align}
\label{eq:appendix-unicyclic-2}
     d_{\epsilon_u}j_e &= P_{11}\beta_1\partial_{\Psi_1}j_1 -P_{44}\beta_1\partial_{\Psi_4}j_4\,, \\
     d_{\epsilon_d}j_e &= -P_{22}\beta_2\partial_{\Psi_2}j_2 + P_{33}\beta_2\partial_{\Psi_3}j_3\,,
\end{align}
\end{subequations}
To proceed we notice that
\begin{align}
\label{eq:appendix-unicyclic-1}
    \partial_{\Psi_e}j_e = -\Gamma_e\Big[\frac{\pi_{s(+e)}\exp(\Psi_e)}{(1+\exp(\Psi_e))^2}+\frac{\pi_{s(-e)}\exp(-\Psi_e)}{(1+\exp(-\Psi_e))^2}\Big]\leq 0\,,
\end{align}
which with the bounds $0 \leq P_{ee}\leq 1$ give us
\begin{subequations}
\begin{align}
\label{eq:appendix-unicyclic-2}
     -|\partial_{\Psi_1}j_1| \leq T_1 d_{\epsilon_u}j_e &\leq |\partial_{\Psi_4}j_4|
     \,, \\
     -|\partial_{\Psi_3}j_3|\leq T_2 d_{\epsilon_d}j_e &\leq |\partial_{\Psi_2}j_2|\,.
\end{align}
\end{subequations}
We want to note that if the response $T_1 d_{\epsilon_u}j_e$ (resp. $T_2 d_{\epsilon_d}j_e$) is saturated, then we have $T_2 d_{\epsilon_d}j_e=0$ (resp. $T_1 d_{\epsilon_u}j_e=0$), due to $\sum_e P_{ee}=1$. Similarly, one can find bounds for the other parameters $\beta_i$, $\mu_i$.

\section{Independent diagonal elements}
\label{sec:blocks}

\begin{figure}
    \centering
    \includegraphics[width=0.4\textwidth]{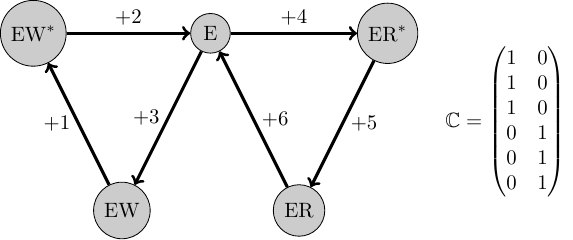}
    \caption{Model of proofreading \cite{murugan2014discriminatory}. The enzyme $\text{E}$ is involved in two complex reactions with wrong $\text{EW}^*$, $\text{EW}$ and right $\text{ER}^*$, $\text{ER}$ species.  
    The matrix $\mathbb{C}$ is block diagonal and vectors $\boldsymbol{c}^\gamma$ are orthogonal.
    }
    \label{fig:isolated}
\end{figure}

Here, we show that $\#\text{ide}$ is defined by the topology of the cycles in the graph $\mathcal{G}$. Using row and column permutations, the matrix $\mathbb{C}$ can be deduced from the block diagonal shape:
\begin{align}
\label{eq:matCblocks}
    \mathbb{C}=\text{diag}\big(\dots,\mathbb{C}_i,\dots\big)\,,
\end{align}
where $\mathbb{C}_i$ are full column rank matrices and all other elements except those matrices  are zero. The examples in the main text have only one block $\mathbb{C}_1=\mathbb{C}$. The blocks in \cref{eq:matCblocks} correspond to the cycles without common edges; see \cref{fig:isolated}. We prove that every diagonal block in \cref{eq:matCblocks} adds one constraint to the diagonal elements, or equivalently $\#\text{ide}=N_e-N_b$, where $N_b$ is the number of blocks $\mathbb{C}_i$ (disconnected cycles in $\mathcal{G}$). Consider the block $\mathbb{C}_i$ of size $n_i\times \gamma_i$. 
We use the decomposition $\mathbb{P}=\mathbb{C}\mathbb{B}$ for block $\mathbb{C}_i$ as $\mathbb{P}_i=\mathbb{C}_i\mathbb{B}_i$, where $\mathbb{B}_i$ and $\mathbb{P}_i$ are diagonal blocks of the matrices $\mathbb{B}$ and $\mathbb{P}$, respectively.  
In addition, we can use the condition $\mathbb{B}\mathbb{C}=\mathbb{I}$ in block form as $\mathbb{B}_i\mathbb{C}_i=\mathbb{I}_{\gamma_i}$, where $\mathbb{I}_{\gamma_i}$ is the identity matrix of size $\gamma_i$. Calculating the trace of the block $\mathbb{P}_i$, we find
\begin{align}
    \tr~\mathbb{P}_i= \tr~\mathbb{C}_i\mathbb{B}_i = \tr~\mathbb{B}_i\mathbb{C}_i=\tr~\mathbb{I}_{\gamma_i}=\gamma_i\,,
\end{align}
which implies the linear constraint on the diagonal elements of $\mathbb{P}$, which consist of the block $\mathbb{P}_i$.

\section{Close to equilibrium.}
Here the goal is to consider Eq.~(25) close to equilibrium.
Local detailed balance (23) implies that at equilibrium, when $\mathcal{F}_e=0$, the steady state probability satisfies detailed balance 
\begin{equation}
\label{DetBal}
\frac{W_{+e}^\text{eq}}{W_{-e}^\text{eq}} =\frac{\pi^\text{eq}_{s(-e)}}{\pi^\text{eq}_{s(+e)}}
\end{equation}
where
\begin{equation}
\label{Gibbs}
\pi^\text{eq}_i = \frac{\exp{(E_i)}}{\sum_{j\in\mathcal{S}}\exp{(E_j)}}\;,
\end{equation}
and as a result all edge currents vanish $j^\text{eq}_e=0$.
The local detailed balance (23) can therefore be rewritten as 
\begin{align}
\label{eq:equilibrium-LDB}
\frac{W_{+e}}{W_{-e}}= \frac{\pi^\text{eq}_{s(-e)}}{\pi^\text{eq}_{s(+e)}} \exp{(\mathcal{F}_e)}\,.
\end{align}
Taking the derivative with respect to $\mathcal{F}_e$, we find that 
\begin{align}
\label{eq:equilibrium-LDB_derv}
\partial_{\mathcal{F}_e} W_{+e}=\Big(\frac{\partial_{\mathcal{F}_e} W_{-e}}{W_{-e}}+1\Big)W_{+e}\;.
\end{align}
Inserting in Eq.~(7), taking the equilibrium limit $\mathcal{F}_e \to 0$, using \cref{DetBal} and introducing the equilibrium traffic $\tau^\text{eq}_{e}$, we find that 
\begin{subequations}
    \label{eq:equilibrium-J}
\begin{align}
 J_{ee}^\text{eq}&=\pi^\text{eq}_{s(+e)}
 \Bigg(\frac{\pi_{s(-e)}^\text{eq}}{\pi_{s(+e))}^\text{eq}}\partial_{\mathcal{F}_e} W_{-e}+W_{+e}^\text{eq}\Bigg)
 -\pi_{s(-e)}^\text{eq}\partial_{\mathcal{F}_e}W_{-e}\nonumber\\
    &=W_{+e}^\text{eq}\pi_{s(+e)}^\text{eq}=\frac{\tau^\text{eq}_{e}}{2}\,,\\
    \mathbb{J}_\text{eq}&=\text{diag}\big(\dots,\frac{\tau^\text{eq}}{2},\dots\big)\,,
\end{align}
\end{subequations}
which defines the equilibrium limit of the Jacobian in Eq.~(25). 
Since by definition
\begin{align}
\label{eq:equilibrium-1}
    \Gamma_{ei}\pi_i &= 
    \left\{
    \begin{array}{cc}
        W_{+e}\pi_{s(+e)}\,,&i=s(+e)  \\
         -W_{-e}\pi_{s(-e)}\,,&i=s(-e)  \\
         0\,,&\text{otherwise}
    \end{array}
    \right. \;,
\end{align}
using \cref{eq:equilibrium-J}, we find that at equilibrium $\Gamma_{ei}^\text{eq}\pi_i^\text{eq} =-S_{ie}J^\text{eq}_{ee}$ and the elements of the reduced matrix $\hat{\mathbb{\Gamma}}_\text{eq}$ can be written as
\begin{align}
\label{eq:equilibrium-2}
    \hat{\Gamma}_{ei}^\text{eq}=-J_{ee}^\text{eq}\Big(\frac{S_{ie}}{\pi_i^\text{eq}}-\frac{S_{Ne}}{\pi_N^\text{eq}}\Big)\,.
\end{align}
We note that \cref{eq:equilibrium-2} implies that $\sum_e\hat{\Gamma}_{ei}^\text{eq}(J_{ee}^\text{eq})^{-1} c^\gamma_e=0$, where $c_e^\gamma$ are the elements of the cycle $\boldsymbol{c}^\gamma$. Therefore, in matrix form, we have:
\begin{align}
\label{eq:equilibrium-3}
\hat{\mathbb{\Gamma}}^\intercal_\text{eq}&\mathbb{J}_\text{eq}^{-1}=\mathbb{A}\hat{\mathbb{S}}\,,\nonumber\\
    \hat{\mathbb{\Gamma}}_\text{eq} &= (\mathbb{A}\hat{\mathbb{S}}\mathbb{J}_\text{eq})^\intercal\,,
\end{align}
where $\mathbb{A}$ is the squared non-singular matrix of free parameters. Indeed, the matrix $\mathbb{A}$ has size $(N_s-1 )\times (N_s-1)$ with the rank $\text{rk}~\mathbb{A}=\text{rk}~\hat{\mathbb{\Gamma}}=N_s-1$, that implies $\det\mathbb{A}\neq 0$. Inserting \cref{eq:equilibrium-3} into Eq.~(8) for the projection matrix, we arrive at
\begin{align}
\label{eq:equilibrium-4}
    \mathbb{P}^\text{eq} &= \mathbb{I}-\mathbb{J}_\text{eq}\hat{\mathbb{S}}^\intercal\mathbb{A}^\intercal(\hat{\mathbb{S}}\mathbb{J}_\text{eq}\hat{\mathbb{S}}^\intercal\mathbb{A}^\intercal)^{-1}\hat{\mathbb{S}}\nonumber\\
    &= \mathbb{I}-\mathbb{J}_\text{eq}\hat{\mathbb{S}}^\intercal(\hat{\mathbb{S}}\mathbb{J}_\text{eq}\hat{\mathbb{S}}^\intercal)^{-1}\hat{\mathbb{S}}\,.
\end{align}
Using \cref{eq:equilibrium-4}, the response matrix (25) becomes
\begin{align}\mathbb{O}=\mathbb{R}^I_{\text{eq}}=\mathbb{X}^\intercal(\mathbb{J}_\text{eq}-\mathbb{J}_\text{eq}\hat{\mathbb{S}}^\intercal(\hat{\mathbb{S}}\mathbb{J}_\text{eq}\hat{\mathbb{S}}^\intercal)^{-1}\hat{\mathbb{S}}\mathbb{J}_\text{eq})\mathbb{X}\,,
\end{align}
which is the semi-positive definite symmetric Onsager matrix of nonequlibrium thermodynamics \cite{Groot1984,forastiere2022linear}.

\section{Calculations for Fig.~3a}
Here we identify the subset of scaled responses marked as disks in Fig.~3a. This subset is defined by the elements of $\mathbb{P}$
\begin{align}
\label{eq:matP-bounded-elements}
    \mathcal{L}(\mathbb{P})=\{P_{ee'}:~P_{ee'}=\sum_{\rho}\nu^{ee'}_\rho P_{\rho\rho} \}\,,
\end{align}
where all $P_{ee'}\in\mathcal{L}(\mathbb{P})$ can be expressed as the linear combinations of the diagonal elements.
QDs model from Fig.~1 exhibits three fundamental cycles (see the graph in Fig.~1b). The topology of this model is defined by:
\begin{align}
\footnotesize
    \mathbb{S}=
    \begin{pmatrix}
-1 & -1 & 0 & 0 & -1 & 0 \\
 1 & 0 & -1 & 0 & 0 & -1 \\
 0 & 1 & 0 & -1 & 1 & 0 \\
 0 & 0 & 1 & 1 & 0 & 1 
    \end{pmatrix}\,,
\overline{\mathbb{C}}=
\begin{pmatrix}
1 & 0 & 0 \\
 0 & 1 & 0 \\
 0 & 0 & 1 
\end{pmatrix}\,,
\Tilde{\mathbb{C}}=
\begin{pmatrix}
 -1 & 0 & 0 \\
 -1 & -1 & 0 \\
 1 & 0 & -1 
\end{pmatrix}\,.\nonumber
\end{align}
which corresponds to $\#\text{ide}=5$.
Inserting $\mathbb{S}$, $\mathbb{C}$  into Eq.~(22) we find 5 independent equations, for example,
\begin{subequations}
\label{eq:tilde-b-example}
\begin{align}
1-(-\Tilde{B}_{22})&=P_{22}\,,\\
1-(-\Tilde{B}_{33})&=P_{33}\,,\\
-\Tilde{B}_{11} &= P_{44}\,, \\
-\Tilde{B}_{12}-\Tilde{B}_{22} &= P_{55}\,,\\
\Tilde{B}_{13}-\Tilde{B}_{33} &= P_{66}\,,
\end{align}
\end{subequations}
which result in
\begin{align}
    \label{eq:matB-solution-3}
    \Tilde{\mathbb{B}}=
    \begin{pmatrix}
        -P_{44} & 1 - P_{55} - P_{22} & P_{66}+P_{33}-1 \\
        b_1 & P_{22}-1 & b_2 \\
        b_3 & b_4 &  P_{33}-1 
    \end{pmatrix}\,,
\end{align}
where $b_1=\Tilde{B}_{21}$, $b_2=\Tilde{B}_{23}$, $b_3=\Tilde{B}_{31}$, $b_4=\Tilde{B}_{32}$ are free unbounded parameters. 
Using \cref{eq:matB-solution-3} we calculate $\mathbb{P}=\mathbb{C}\big(\mathbb{I}-\Tilde{\mathbb{B}}\Tilde{\mathbb{C}},\Tilde{\mathbb{B}}\big)$ as
\begin{widetext}
\begin{align}
\label{eq:matP-example}
\mathbb{P}=
\begin{pmatrix}
 P_{11} & -P_{22}-P_{55}+1 & P_{33}+P_{66}-1 & -P_{44} &
   -P_{22}-P_{55}+1 & P_{33}+P_{66}-1 \\
 b_1-b_2+P_{22}-1 & P_{22} & b_2 & b_1 & P_{22}-1 & b_2 \\
 b_3+b_4-P_{33}+1 & b_4 & P_{33} & b_3 & b_4 & P_{33}-1 \\
 -P_{11} & P_{22}+P_{55}-1 & -P_{33}-P_{66}+1 & P_{44} &
   P_{22}+P_{55}-1 & -P_{33}-P_{66}+1 \\
 -b_1+b_2-P_{11}-P_{22}+1 & P_{55}-1 & -b_2-P_{33}-P_{66}+1 &
   P_{44}-b_1 & P_{55} & -b_2-P_{33}-P_{66}+1 \\
 -b_3-b_4+P_{11}+P_{22}+P_{33}-1 & -b_4-P_{22}-P_{55}+1 & P_{66}-1 &
   -b_3-P_{44} & -b_4-P_{22}-P_{55}+1 & P_{66} \\
\end{pmatrix}\,\,,
\end{align}
\end{widetext}
which shows the bounded elements of $\mathbb{P}$ as those which are expressed in terms of the diagonal elements without $b_k$ with $k=1,2,3,4$.

\section{Physical perturbations: Multicyclic}
Here we illustrate our approach to the perturbations of physical parameters (energy levels) in multicyclic model shown in Fig.~1. Using the expressions for $\Psi_e$ from Fig.~1 with the constant temperature $T_i=T$, we find the Jacobian for $\boldsymbol{p}=(\epsilon_u,\epsilon_d,U_C)$ and calculate the responses [Eq.~6b]
\begin{align}
\label{eq:response-model-parameters}
\big[Td_{p_{e'}}j_e\big]_{e,e'}=\mathbb{P}\cdot
\begin{pmatrix}
        \partial_{\Psi_1}j_1 & 0 & 0 \\
        0 & \partial_{\Psi_2}j_2 & 0\\
        0 & \partial_{\Psi_3}j_3 & \partial_{\Psi_3}j_3 \\
        \partial_{\Psi_4}j_4 & 0 &  \partial_{\Psi_4}j_4\\
        0 &  \partial_{\Psi_5}j_5 & 0\\
        0 &  \partial_{\Psi_6}j_6 & \partial_{\Psi_6}j_6
    \end{pmatrix}\,,
\end{align}
where the derivatives $\partial_{\Psi_e}j_e\leq 0$; see \cref{eq:appendix-unicyclic-1}. 
Using \cref{eq:matP-example}, we find that the responses $j_\alpha$ with $\alpha=1,4$ are bounded to all perturbations of the all edge parameters $\Psi_e$. We exploit it to find the bounds for $T d_p j_1$:
\begin{subequations}
\begin{align}
    T d_{\epsilon_u}j_1 &= P_{11}\partial_{\Psi_1}j_1- P_{44}\partial_{\Psi_4}j_4\,,\\
    \partial_{\Psi_1}j_1 &\leq T d_{\epsilon_u}j_1 \leq |\partial_{\Psi_4}j_4|\,,
\end{align}
\end{subequations}
\begin{subequations}
\begin{align}
 T d_{\epsilon_d}j_1 &= (-P_{22}-P_{55}+1)\partial_{\Psi_2}j_2+(P_{33}+P_{66}-1)\partial_{\Psi_3}j_3 \nonumber\\
 +(-P_{22}&-P_{55}+1)\partial_{\Psi_5}j_5+ 
   (P_{33}+P_{66}-1)\partial_{\Psi_6}j_6\,, \\
|T d_{\epsilon_d}j_1| &\leq |\partial_{\Psi_2}j_2|+|\partial_{\Psi_3}j_3|+|\partial_{\Psi_5}j_5|+|\partial_{\Psi_6}j_6|\,,
\end{align}
\end{subequations}
\begin{subequations}
\begin{align}
 T d_{U_C}j_1 &= (P_{33}+P_{66}-1)\partial_{\Psi_3}j_3 -P_{44}\partial_{\Psi_4}j_4 \nonumber\\
 &  +(P_{33}+P_{66}-1)\partial_{\Psi_6}j_6\,, \\
T d_{U_C}j_1 &\leq |\partial_{\Psi_3}j_3|+|\partial_{\Psi_4}j_4|+|\partial_{\Psi_6}j_6|\,,\\
T d_{U_C}j_1 &\geq-|\partial_{\Psi_2}j_2|-|\partial_{\Psi_6}j_6|\,.
\end{align}
\end{subequations}
In addition, using \cref{eq:matP-example} we have similar expressions for the response $d_{p_e}j_2=-d_{p_e}j_1$.


\bibliography{biblio}
\end{document}